\documentstyle[twoside,fleqn,espcrc2,epsfig]{article}

\newcommand{\AmS}{{\protect\the\textfont2
  A\kern-.1667em\lower.5ex\hbox{M}\kern-.125emS}}

\newcommand{\be}{\begin{equation}}                                              
\newcommand{\ee}{\end{equation}}                                                
\newcommand{\half}{\frac{1}{2}}

\title{Numerical study of SU(2) Yang-Mills theory with gluinos}

\author{G. Koutsoumbas \address{Physics Department, National Technical
                                University, Athens, Greece},
	I. Montvay \address{Deutsches Elektronen Synchrotron, DESY,
                            Notkestr. 85, D-22603 Hamburg, Germany},
	A. Pap \address{Department of Physics, University of
                        Cincinnati, Cincinnati, Ohio 43221},
	K. Spanderen\thanks{Poster presented by K. Spanderen.}
           \address{Institut f\"ur Theoretische Physik I,
           Universit\"at M\"unster, Wilhelm-Klemm-Str. 9, \\
        D-48149 M\"unster, Germany},
        D. Talkenberger\ $\rm^d$ and
        J. Westphalen\ $\rm^b$}       
\begin{document}

\begin{abstract}
 We report on a numerical investigation of the SU(2) gauge theory with
 gluinos.
 The low-lying spectrum in bosonic and fermionic channels is 
 determined.
 Improvements of the multi-bosonic algorithm are discussed.
\end{abstract}

\maketitle

\section{INTRODUCTION}

 Simulating supersymmetric field theory on the lattice is not
 straightforward.
 As the Poincar\'e symmetry is broken on the lattice, so is SUSY.
 A possible way out is to follow the ansatz by Curci and Veneziano
 \cite{CURVEN}.
 They proposed to live with broken SUSY on the lattice and to reveal 
 the relevant information in the continuum limit, where SUSY should be
 restored.

\section{ACTION AND ALGORITHMS}

 The {\em Curci-Veneziano action} of $N=1$ SYM, after integrating out
 the gluino field, is given by
\be\label{eq01}
S_{CV} = \beta\sum_{pl} \left( 1-\half{\rm Tr\,}U_{pl} \right)
- \half\log\det Q[U] \ .
\ee
 Here the fermion matrix is
\be\label{eq02}
Q_{yv,xu} = \delta_{yx}\delta_{vu} - 
K\sum_{\mu=\pm} \delta_{y,x+\hat{\mu}}(1+\gamma_\mu) V_{vu,x\mu} 
\ee
 and $V_{vu,x\mu}$ the gauge link variable in the adjoint (triplet)
 representation.
 The bare parameters are: the usual gauge coupling $\beta=4g^{-2}$ and
 the hopping parameter of gluino $K$.
 The factor of $\half$ in front of $\log\det$ shows that the gluino is
 a Majorana fermion, corresponding effectively to a flavour number
 $N_f=\half$.
 (For further discussion of the lattice action see \cite{PLENAR}.)

\subsection{Two-step multi-bosonic algorithm}

 For the Monte Carlo simulations with dynamical gluinos we use the
 two-step variant \cite{GLUINO} of the multi-bosonic algorithm
 \cite{LUSCHER}.
 This requires smaller storage and has shorter autocorrelations.
 For the correction step with the higher order polynomial $P_n$ it
 turned out better to use another method, not the one introduced in
 ref.~\cite{GLUINO}.
 In terms of the hermitean fermion matrix
 $\tilde{Q}\equiv\gamma_5 Q=\tilde{Q}^\dagger$ let us define, with $n$
 even, the decomposition
\begin{eqnarray}\nonumber
P_n(\tilde{Q}^2) & = & r_0 \prod_{j=1}^{n/2}
(\tilde{Q}-\rho_j^*)(\tilde{Q}-\rho_j)
\\
 & \equiv & P_{n/2}(\tilde{Q})^\dagger P_{n/2}(\tilde{Q}) \ .
\end{eqnarray}
 Using this form, the noise vector $\eta$, necessary in the {\em noisy
 correction} step, can be generated from the gaussian vector
 $\eta^\prime$ according to
\be\label{eq04}
\eta=P_{n/2}(\tilde{Q})^{-1}\eta^\prime \ ,
\ee
 where $P_{n/2}(\tilde{Q})^{-1}$ can be obtained as
\be\label{eq05}
P_{n/2}(\tilde{Q})^{-1}=\frac{P_{n/2}(\tilde{Q})^\dagger}
{P_n(\tilde{Q}^2)} \simeq 
P_{-n}(\tilde{Q}^2) P_{n/2}(\tilde{Q})^\dagger \ .
\ee
 In the last step $P_{-n}$ denotes a polynomial approximation for
 the inverse of $P_n$ on the interval $[\epsilon,\lambda]$, which
 covers the spectrum of $\tilde{Q}^2=Q^\dagger Q$ on typical gauge
 configurations.
 For the calculation of the necessary polynomials procedures written
 in Maple are available \cite{POLYNOM}.

\subsection{Preconditioning and eigenvalue distributions}

\begin{figure}[th]
\vspace*{-0.9cm}
\begin{center}
\epsfig{file=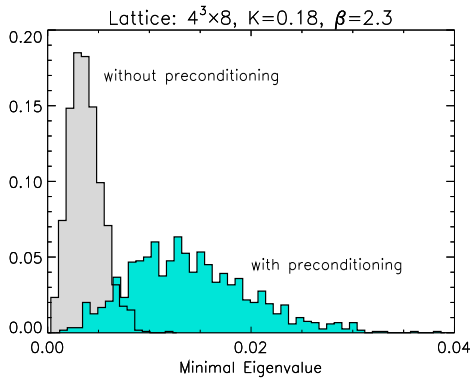,
        width=7cm,height=5.25cm}
\vspace*{-1.1cm}
\caption{\label{fig01}
 Distribution of the minimal eigenvalues of $\hat{Q}^2$ and
 $\tilde{Q}^2$ on $4^3 \cdot 8$ lattice at ($\beta=2.3$, $K=0.18$).}
\end{center}
\vspace*{-0.9cm}
\end{figure}

 In order to improve the performance of our fermion simulation
 algorithm, preconditioning according to ref.~\cite{JEGERL} turned out
 to be very useful.
 The hermitean fermion matrix is decomposed as
\be\label{eq06}
\tilde{Q} = \gamma_5 Q = \left(
\begin{array}{cc}
\gamma_5  &  -K\gamma_5 M_{oe}  \\  -K\gamma_5 M_{eo}  &  \gamma_5
\end{array} \right)
\ee
 and then we have
\be\label{eq07}
\det\tilde{Q} = \det\hat{Q} \ ,
\;\;{\rm with}\;\;
\hat{Q} \equiv \gamma_5 - K^2\gamma_5 M_{oe}M_{eo} \ .
\ee
 The matrix {$\hat{Q}^2$} has a smaller condition number
 {$\lambda/\epsilon$}  than {$\tilde{Q}^2$}.
 The condition number and its fluctuations on different gauge
 configurations are dominated by the minimal eigenvalue.
 A comparison of the fluctuations of the lowest eigenvalue of
 $\hat{Q}^2$ and $\tilde{Q}^2$ is shown in figure~\ref{fig01}.

 As a consequence of the smaller condition numbers and smaller orders 
 of the polynomial approximations the autocorrelations are shorter
 with proconditioning than without it.
 For an example see table~\ref{tab01}.
\begin{table}[bh]
\vspace*{-0.9cm}
\begin{center}
\caption{\label{tab01}
 Autocorrelation times for the plaquette value on $4^3 \cdot 8$
 lattice at $(\beta=2.3, K=0.18)$.
 $\delta^2$ is the deviation norm of the polynomial approximation.}
\vspace*{0.2cm}
\begin{tabular}{|c|c|c|c|c|}\hline
$n$ & $\varepsilon$ & $\delta^2$ & precond. & $\tau_{int}( \Box )$ \\ 
\hline\hline
16 & 0.0004 & $0.00085$  & & 194(38) \\ \hline
8 & 0.002 & $0.00052$ & $\times$ & 65(17) \\ \hline\hline
\end{tabular}
\end{center}
\vspace*{-0.3cm}
\end{table}

\begin{figure}[t]
\vspace*{-0.3cm}
\begin{center}
\epsfig{file=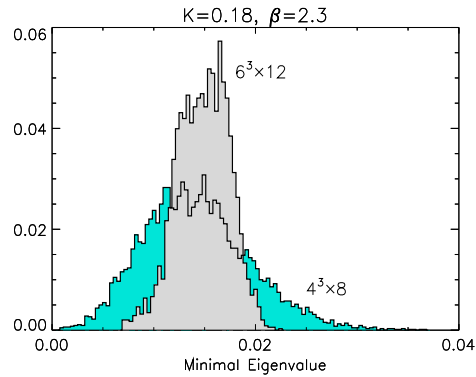,
        width=7cm,height=5.25cm}
\vspace*{-1.0cm}
\caption{\label{fig02}
 Distribution of minimal eigenvalues of the squared preconditioned
 hermitean fermion on $4^3 \cdot 8$ and $6^3 \cdot 12$ lattice at
 $(\beta=2.3, K=0.18)$.}
\end{center}
\vspace*{-1.1cm}
\end{figure}

 Random matrix models suggest that the fluctuations of the minimal
 eigenvalue are inversely proportional to the lattice volume.
 (For references and a recent summary see ref.~\cite{VERBAAR}.)
 This is advantageous for the choice of the interval of polynomial
 approximations $[\epsilon,\lambda]$.
 Our numerical data support the decrease of fluctuations, as is shown by
 figure~\ref{fig02}.

\section{LOW-LYING SPECTRUM}
 
 A first important question for the numerical simulations is to
 determine the masses of lightest states in different channels as
 a function of the gluino mass.
 These can be compared to the predictions of the low energy effective
 action \cite{VENYAN},\cite{MASVEN}.
 It is expected that there is confinement as in pure gauge theory.
 In the limit of zero gluino mass the states should occur in degenerate
 supermultiplets.

 The methods to determine low-lying masses can be tested and tuned on
 ``quenched'' gauge configurations generated with the pure gauge part
 of the action \cite{KOUMON},\cite{DOGUHEVL}.

\subsection{Correlations}

\begin{figure}[t]
\begin{center}
\vspace*{-0.6cm}
\epsfig{file=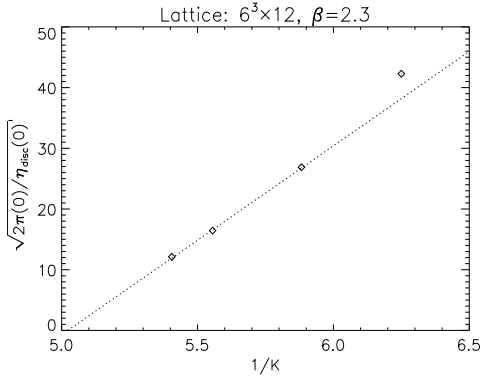,
        width=7cm,height=5.25cm}
\vspace*{-1.2cm}
\caption{\label{fig03}
 The square root of the ratio of the connected to disconnected part of
 the correlation for a-eta as a function of $1/K$.
 The line is a linear extrapolation through the last three points.}
\end{center}
\vspace*{-1.0cm}
\end{figure}

 The lightest supermultiplet at zero gluino mass is presumably a
 massive chiral multiplet consisting of a scalar, a pseudoscalar and
 a spin-$\half$ Majorana fermion.
 The bosonic states have analogues in QCD and can be made out of
 gluinos.
 Since the gluinos are in the adjoint representation, let us call them
 {\em a-eta} and {\em a-$f_0$} for pseudoscalar and scalar,
 respectively.
 The states made out of gluinos can be generally called
 {\em gluinoballs}.
 The two-fermion correlation functions contain connected and 
 disconnected parts.
 The connected parts can be considered separately and, in analogy with
 QCD, can be associated with the {\em a-pion} and {\em a-sigma} for
 pseudoscalar and scalar, respectively.
 Near the {\em critical hopping parameter} $K_{cr}$ corresponding to
 zero gluino mass the ratio of the connected to disconnected part has
 to be ${\cal O}(1)$.
 This is how the dependence of the chiral symmetry breaking pattern
 on the number of flavours can become manifest.
 As our numerical data obtained with the {\em volume source method}
 \cite{KFMOU} show, for $K < K_{cr}$ the disconnected part is
 much smaller, but it is increasing for increasing $K$ (see
 figure~\ref{fig03}).
 Extrapolation to a ratio equal to one gives a rough estimate
 $K_{cr} \simeq 0.20$.

\subsection{Masses}

 Besides the gluinoballs, confined states can also be composites of
 gluons and gluinos.
 Purely gluonic states are, as usual, called {\em glueballs} and 
 the spin-$\half$ fermionic states made out of gluinos and gluons as
 {\em gluino-glueballs}.

\begin{figure}[t]
\begin{center}
\vspace*{-0.6cm}
\epsfig{file=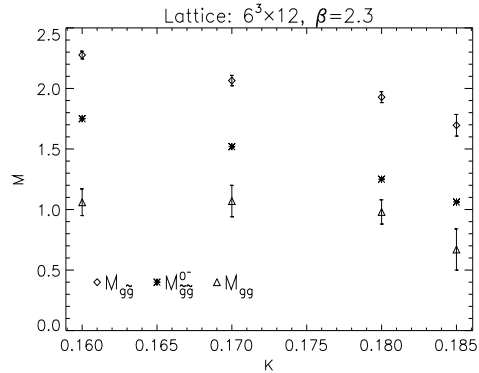,width=7cm,height=5.25cm}
\vspace*{-1.2cm}
\caption{\label{fig04}
 Dependence of the lowest bound state masses on the hopping parameter
 $K$ on $6^3 \cdot 12$ lattices at $\beta=2.3$.}
\end{center}
\vspace*{-0.8cm}
\end{figure}

 Our first series of Monte Carlo runs with dynamical gluinos has been
 performed at $\beta=2.3$ and $0.16 \leq K \leq 0.185$ on $4^3 \cdot 8$
 and $6^3 \cdot 12$ lattices.
 For the determination of the masses of glueballs and gluino-glueballs
 we used smeared sources \cite{TEPER}.
 Preliminary results on the masses of $0^+$ glueball, a-eta and
 spin-$\half$ gluino-glueball are shown in figure~\ref{fig04}.
 As one can see, in this range of couplings the expected degeneracy of
 bound state masses is not yet observed.
 We are presently extending the runs to other parameter values and
 to larger lattices. 

\raggedbottom


\end{document}